\documentclass[11pt]{article}
\usepackage[colorlinks=true, citecolor=blue, urlcolor=blue, linkcolor=blue, breaklinks=true, pdfpagelabels=false]{hyperref}
\usepackage{cite}

\usepackage{hyperref}

\usepackage{amsmath,amsfonts,amssymb}
\usepackage{graphicx}
\usepackage{epsfig,epsf}
\usepackage{bm}
\usepackage{xcolor}

\def\a {\alpha}
\def\b {\beta}

\def\l {\lambda}

\def\bar {\overline}

\def\be {\begin{equation}}
\def\ee {\end{equation}}
\def\bea {\begin{eqnarray}}
\def\eea {\end{eqnarray}}
\def\n {\nonumber}

\def\barr{\begin{array}}
\def\earr{\end{array}}

\def\opcit(#1){ {\em op. cit.}, #1}

\def\issue(#1,#2,#3){#1, #2 (#3)} 

\def\equationautorefname~#1\null{Eq.\,(#1)\null}
\def\pageautorefname\nobreakspace{p.}

\makeatletter\renewcommand{\p@subsection}{\thesection.}\makeatother%
 
\begin{document}

\renewcommand*{\thefootnote}{\fnsymbol{footnote}}


\begin{center}
{\Large\bf{Light Higgs in the Georgi Machacek Model : neutral or charged?}}


\vspace{5mm}

{\bf Swagata Ghosh}$^{a}$\footnote{swgtghsh54@gmail.com}

\vspace{3mm}
{\em{${}^a$\ \ \ Department of Physics, Indian Institute of Technology Kharagpur, Kharagpur 721302, India.}}

\end{center}

\begin{abstract}
The ATLAS and CMS experiments at the LHC, as well as the LEP and Tevatron, observed the existence of a light SM-like Higgs boson with mass about 95 GeV. 
These experiments also record the evidences of a light charged Higgs boson with mass less than the top-quark mass. 
The Georgi-Machacek (GM) model consists of six charged Higgs, one CP-odd and one CP-even Higgs besides the SM Higgs and the SM-like Higgs bosons. 
Considering the theoretical constraints along with the $\sqrt{s} = 13$ TeV Higgs signal strengths, this paper shows that the presence of the 95 GeV SM-like Higgs boson disfavours the presence of the other eight low-mass (below $160$ GeV) Higgs. 
This also triggers a high triplet vev above $40$ GeV. 

\end{abstract}



\setcounter{footnote}{0}
\renewcommand*{\thefootnote}{\arabic{footnote}}

\section{Introduction}
\label{intro}
%

After the discovery of the Higgs boson of the Standard Model (SM) with the mass about $125$ GeV \cite{ATLAS:2012yve,CMS:2012qbp}, the LHC devote searches for other SM-like scalars with mass higher or lower than the SM Higgs boson. 
The search for the Higgs at the LHC are still unable to confirm the absence of other scalar resonances. 
This leads to the existence of many beyond SM (BSM) models through the extension of the scalar sector of the Standard Model. 
One of the neutral scalars of these models must be the SM Higgs boson. 
The other Higgs, whether they are scalars or pseudoscalars, can be neutral as well as charged, with masses greater or less than $125$ GeV. 

The search for the $SM-like$ Higgs boson with mass below $125$ GeV was started at the LEP \cite{OPAL:2002ifx,LEPWorkingGroupforHiggsbosonsearches:2003ing,ALEPH:2006tnd} and the Tevatron \cite{CDF:2012wzv} much before the similar searches at the LHC \cite{CMS:2015ocq,CMS:2018cyk,CMS:2018rmh,ATLAS:2018xad,CMS:2022goy,ATLAS:2022abz,CMS:2023yay}. 
The searches of CMS at $\sqrt{s} = 8$ TeV and $13$ TeV showed an excess around $95.3$ GeV \cite{CMS:2015ocq,CMS:2018cyk} with local significance of $2.8 \sigma$. 
Subsequently, another publication from CMS \cite{CMS:2023yay} expressed the excess around $95.4$ GeV, provided in terms of the signal strength of Higgs to di-photon decay. 
ATLAS \cite{ATLAS:2018xad} also reported the local excess of $1.7\sigma$ for the light Higgs with mass about $95.4$ GeV to di-photon signal strength. 
The combined signal strength for this channel around the same mass is also available in the literature \cite{Biekotter:2023oen} with an excess of $3.2\sigma$. 
For the decay of the light Higgs to di-tau, the signal strength with a local excess of $2.2\sigma$ is observed by CMS \cite{CMS:2022goy} around the mass of $95$ GeV, but no such excess is published by ATLAS till date. 
LEP \cite{LEPWorkingGroupforHiggsbosonsearches:2003ing} observed a local significance of $2.3\sigma$ from the signal strength of the low mass Higgs around $95.4$ GeV to $b\bar{b}$ decay. 
Considering the above findings, one can always inspect for a light SM-like Higgs with mass around $95.4$ GeV in different BSM models \cite{Fox:2017uwr,Haisch:2017gql,Biekotter:2019kde,Aguilar-Saavedra:2020wrj, Ahriche:2023wkj, Chen:2023bqr}. 

Besides the low mass SM-like Higgs bosons, CMS and ATLAS at the LHC also search for charged scalars or pseudoscalars whether it is lighter or heavier than the top-quark. 
As this work focuses only on the low mass resonances, search for charged Higgs bosons with mass less than $m_t$ are considered. 
The inspection of the light charged resonances at the LHC include its decay to $cb$ \cite{Ivina:2022tfm, ATLAS:2021zyv}, $cs$ \cite{CMS:2020osd, ATLAS:2024oqu, ATLAS:2013uxj, ATLAS:2010ofa}, and $\tau\nu_{\tau}$ \cite{CMS:2019bfg, Abbaspour:2018ysj, ATLAS:2016avi, ATLAS:2011pka, ATLAS:2012nhc, Ali:2011qf, CMS:2012fgz}. 
The results of these decay channels are also explored in different BSM models \cite{Akeroyd:2018axd, Arhrib:2020tqk, Chakraborti:2021bpy, Jueid:2021avn, Arhrib:2021xmc, Benbrik:2021wyl, Cheung:2022ndq, Akeroyd:2022ouy, Slabospitskii:2022asy, Hu:2022gwd, Kim:2022nmm, Arhrib:2024sfg, Duarte:2024zeh, Arroyo-Urena:2024soo, Kim:2024jhx, Bae:2024lov, Ghosh:2022wbe}. 

This work pays attention to the simultaneous presence of the low mass SM-like Higgs boson as well as low mass charged Higgs boson. 
For this, we consider the triplet extension of the SM, where the electroweak $\rho$ - parameter is unity at the tree level $i.e.$, the Georgi-Machacek (GM) model \cite{Georgi:1985nv}. 
This novel characteristic of the GM model allows the triplet vacuum expectation value (vev) $v_2$ to be large enough, unlike the other triplet extensions of the SM. 
The extended scalar sector here incorporates two $SU(2)_L$ triplets, one real and one complex, retaining the same vev. 
A total number of ten Higgs bosons containing two CP-even neutral scalars ($h, S$), one neutral pseudoscalar ($H_3^0$), one neutral fermiophobic scalar ($H_5^0$), two singly charged scalars ($H_5^{\pm}$), two singly charged pseudoscalars ($H_3^{\pm}$), two doubly charged scalars ($H_5^{\pm\pm}$) are present in this model. 
The SM Higgs boson can be identified as $h$, and $S$ is the SM-like Higgs boson. 
The mass of $h$ ($m_h$) can be greater or less than the mass of $S$ ($m_S$). 
Here as this work focuses on the light Higgs, we consider $m_S < m_h$. 
In GM model, $H_3^{0,\pm}$ and $H_5^{0,\pm,\pm\pm}$ share the common masses $m_3$ and $m_5$ respectively. 
Being fermiophobic, the singly charged scalars  $H_5^{\pm}$ do not couple to the fermions, but the other two singly charged Higgs bosons $H_3^{\pm}$ have couplings with the fermions. 
The emphasis on the light charged Higgs boson obligates $m_3 < m_t$, where $m_t$ is the top-quark mass. 
The theoretical constraints of the model \cite{Ismail:2020zoz, Hartling:2014zca, Krauss:2017xpj, Moultaka:2020dmb, Aoki:2007ah} at the same time as the LHC Higgs signal strengths probe the corresponding values of $m_5$, such that $m_5$ is not too far from $m_3$. 
The global fits \cite{Chiang:2018cgb} as well as the indirect constraints \cite{Hartling:2014aga} also probe the available values of the parameters.  
A calculator \cite{Hartling:2014xma} is also available to scan the parameter space. 

The presence of $95$ GeV Higgs boson in the context of the GM model is studied before \cite{Ahriche:2023wkj, Chen:2023bqr} but without considering the bounds on the low triplet mass \cite{Ghosh:2022wbe}. 
This work considers that constraints together with the other constraints and shows that the presence of a $95$ GeV SM-like Higgs boson disallows the presence of other low mass Higgs bosons. 
In other words, the presence of low mass charged Higgs bosons disfavours the $95$ GeV SM-like Higgs boson. 
The section \ref{model} describes the Georgi-Machacek model in brief with the constraints in the section \ref{constraints}. 
The section \ref{Results} gives the results. 
The conclusions are given in the section \ref{Conclusions}.

 \section{The Georgi-Machacek Model}
 \label{model}
%
The scalar sector of the Georgi-Machacek model contains the SM doublet, one real triplet and one complex triplet \cite{Hartling:2014zca}. 
This doublet can be written in terms of a bi-doublet $\Phi$ and the triplets can be written in terms of a bi-triplet $X$. 
The scalar potential of the GM model as the function of $\Phi$ and $X$ is given by, 
 \bea
 V\left(\Phi,X\right) &=& \frac{{\mu_2}^2}{2}\, {\rm Tr}\left(\Phi^\dag\Phi\right) 
 + \frac{{\mu_3}^2}{2}\, \rm{Tr}\left(X^\dag X\right)
 + {\l_1}\left[{\rm Tr}\left(\Phi^\dag\Phi\right)\right]^2  
 + {\l_2}\, {\rm Tr}\left(\Phi^\dag\Phi\right)\, {\rm Tr}\left(X^\dag X\right) \n\\
 && + {\l_3}\, {\rm Tr}\left(X^\dag X X^\dag X\right)
  + {\l_4}\left[{\rm Tr}\left(X^\dag X\right)\right]^2 
 - \frac{{\l_5}}{2}\, {\rm Tr}\left(\Phi^\dag \sigma^a \Phi\tau^b\right)\, {\rm Tr}\left(X^\dag t^a X t^b\right) \n\\
 && - \frac{{M_1}}{2} \, {\rm Tr}\left(\Phi^\dag \sigma^a \Phi\tau^b\right) {\left(U X U^\dag\right)_{ab}} 
 - {M_2} \, {\rm Tr}\left(X^\dag t^a X t^b\right) {\left(U X U^\dag\right)_{ab}}\, .
 \label{eq:genPot} 
 \eea
Here the three Pauli matrices are given by $\sigma^a$. 
The matrices $t^a$s and $U$ are as follows,
\be
t^1=\frac{1}{\sqrt2}
\begin{pmatrix}
 0 & 1 & 0\cr
 1 & 0 & 1\cr
 0 & 1 & 0
\end{pmatrix}
\,,
t^2=\frac{1}{\sqrt2}
\begin{pmatrix}
 0 & -i & 0\cr
 i & 0 & -i\cr
 0 & i & 0
\end{pmatrix}
\,,
t^3=
\begin{pmatrix}
 1 & 0 & 0\cr
 0 & 0 & 0\cr
 0 & 0 & -1
\end{pmatrix}\,,
U=
\frac{1}{\sqrt{2}}\begin{pmatrix}
 -1 & 0 & 1\cr
 -i & 0 & -i \cr
 0 & \sqrt{2} & 0
\end{pmatrix}\,.
\ee
The $SU(2)_L$ real ($Y=0$) and complex ($Y = 2$) triplet, $\left(\xi^{+},\xi^0,\xi^{-}\right)^T$ and $\left(\chi^{++},\chi^{+},\chi^0\right)^T$ respectively, together form the bi-triplet 
\be
X =
\begin{pmatrix}
 \chi^{0*} & \xi^{+} & \chi^{++}\cr
 \chi^{-} & \xi^0 & \chi^{+}\cr
 \chi^{--} & \xi^{-} & \chi^0
\end{pmatrix}\,.
\ee
The preservation of the tree-level custodial symmetry, $i.e.$, $\rho_{\rm tree}\, \equiv M_W^2/M_Z^2 \cos^2{\theta_W} = 1$, demands the equality of the two triplet vevs after the electroweak symmetry breaking. 
The relation between the electroweak vev $v$ with the doublet vev $v_1$ and the triplet vev $v_2$ is given by, 
$\sqrt{{v_1}^2+8{v_2}^2}=v\approx246~\text{GeV}\,.
$ 
The physical fields of the GM model are listed below : 
\bea
&&H_5^{++}=\chi^{++}\,, \ \ \ 
H_5^{+}=\frac{\left(\chi^{+}-\xi^{+}\right)}{\sqrt{2}}\,, \ \ \ 
H_5^0=\sqrt{\frac23}\xi^0-\sqrt{\frac13}\chi^{0R}\,, \n\\
&&H_3^{+}=-\sin {\b}\, \phi^{+}+\cos {\b}\, \frac{\left(\chi^{+}+\xi^{+}\right)}{\sqrt{2}}\,, \ \ \ 
H_3^0=-\sin {\b}\, \phi^{0I}+\cos {\b}\, \chi^{0I}\,, \n\\
&&H_1^0=\phi^{0R}\,, \ \ \ 
H_1^{0'}=\sqrt{\frac13}\xi^0+\sqrt{\frac23}\chi^{0R}\,,
\label{eq:fields}
\eea
with 
$\tan {\b} = 2\sqrt{2}{v_2}/v_1\,$. 
The components of the custodial quintuplet ($H_5^{\pm\pm,\,\pm,\,0}$) and the custodial triplet ($H_3^{\pm,\,0}$) share the degenerate masses $m_5$ and $m_3$, respectively. 
The square of these common masses are given by, 
\bea
{m_5}^2&=&\frac{M_1}{4{v_2}}{v_1}^2+12{M_2}{v_2}+\frac32{\l_5}{v_1}^2+8{\l_3}{v_2}^2\,, \n\\
{m_3}^2&=&\left(\frac{M_1}{4{v_2}}+\frac{\l_5}{2}\right)v^2\,.
\label{eq:m3m5}
\eea
The two custodial singlets $H_1^0$ and $H_1^{0^{\prime}}$ mix with each other through the mixing angle $\a$, such that, 
$
\tan{2\a}=(2 {{\cal M}_{12}}^2)/({{\cal M}_{22}}^2-{{\cal M}_{11}}^2)\,.
$ 
Here ${\cal M}_{11,12,22}^2$ are the components of the mass-squared matrix, given by, 
$
{\cal M}_{11}^2 = 8 {\l_1}{v_1}^2
$, 
$
{\cal M}_{12}^2 = {\cal M}_{21}^2 = \frac{\sqrt{3}}{2}\left[-M_1+4\left(2\l_2-\l_5\right)v_2\right]v_1
$, 
$
{\cal M}_{22}^2 = 
\frac{M_1 v_1^2}{4 v_2} - 6 M_2 v_2 + 8 (\lambda_3 + 3 \lambda_4) v_2^2
$.
We have two mass eigenstates $S_{1,2}$. 
Here $S_2$ is always heavier than $S_1$. 
Any one of these two singlets can be considered as the SM Higgs with mass $\sim 125$ GeV. 
The SM Higgs is identified as $h$ and the other one is $S$. 

For $m_h > m_S$,
\be
S = \cos {\a}\,\, H_1^0-\sin {\a}\,\, H_1^{0'}\,, \, 
h = \sin {\a}\,\, H_1^0+\cos {\a}\,\, H_1^{0'}\,,
\label{eq:mhmH}
\ee
with $m_S^2 = \frac12 \left[{{\cal M}_{11}}^2+{{\cal M}_{22}}^2-\sqrt{\left({{\cal M}_{11}}^2-{{\cal M}_{22}}^2\right)^2 +4\left({{\cal M}_{12}}^2\right)^2}\right]\,.
$

For $m_h < m_S$,
\be
h = \cos {\a}\,\, H_1^0-\sin {\a}\,\, H_1^{0'}\,, \, 
S = \sin {\a}\,\, H_1^0+\cos {\a}\,\, H_1^{0'}\,,
\label{eq:mhmH}
\ee
with $m_S^2 = \frac12 \left[{{\cal M}_{11}}^2+{{\cal M}_{22}}^2+\sqrt{\left({{\cal M}_{11}}^2-{{\cal M}_{22}}^2\right)^2 +4\left({{\cal M}_{12}}^2\right)^2}\right]\,.
$
%
 \section{Theoretical constraints and LHC data}
 \label{constraints}
The theoretical constraints limit the values of the quartic couplings $\l_{1-5}$ which result in probing the parameter space of the GM model. 
The electroweak vacuum stability constraints are available in the literature as,
 \bea
 \l_1\, &>&\, 0\,,\n\\
 \l_2+\l_3\,&>&\,0\,,\n\\
 \l_2+\frac12\l_3\,&>&\,0\,,\n\\
 -\mid\l_4\mid +2\sqrt{\l_1\left(\l_2+\l_3\right)}\,&>&\,0\,,\n\\
 \l_4-\frac14 \mid\l_5\mid +\sqrt{2\l_1\left(2\l_2+\l_3\right)}\,&>&\,0\,,
 \label{eq:stability}
 \eea
and the perturbative unitarity constraints are given by,
 \bea
 \sqrt{\left( 6\l_1-7\l_3-11\l_4 \right)^2+36\l_2^2}+\mid 6\l_1+7\l_3+11\l_4\mid &<& 4\pi\,, \n\\
 \sqrt{\left( 2\l_1+\l_3-2\l_4 \right)^2+\l_5^2}+\mid 2\l_1-\l_3+2\l_4\mid &<& 4\pi\,, \n\\
 \mid 2\l_3+\l_4\mid &<& \pi\,, \n\\
 \mid \l_2-\l_5\mid &<& 2\pi\,.
 \label{eq:unitarity}
 \eea
These constraints collectively result in the theoretical constraints \cite{Hartling:2014zca, Hartling:2014aga}. 
The LHC Higgs signal strength data at $\sqrt{s} = 13$ TeV  \cite{CMS:2018lkl,ATLAS:2019slw} also curb the parameter space of the GM model. 

Also, for the search of light charged Higgs boson $H_3^{\pm}$, the experimental results for the channel $p p \rightarrow t \bar{t}, \, t(\bar{t}) \rightarrow H_3^+ (H_3^-) b, \, H_3^+ (H_3^-) \rightarrow c\bar{s} (\bar{c}s) \, {\rm or}\, H_3^+ (H_3^-) \rightarrow \tau^+ \nu_{\tau} (\tau^- \bar{\nu}_{\tau})$ 
from ATLAS \cite{ATLAS:2024oqu, ATLAS:2013uxj, ATLAS:2010ofa, ATLAS:2018gfm} and CMS data \cite{CMS:2020osd,CMS:2019bfg} curb the parameter space heavily \cite{Ghosh:2022wbe}. 
%
%
 \begin{figure}
  \begin{center}
  \includegraphics[width= 7.5cm]{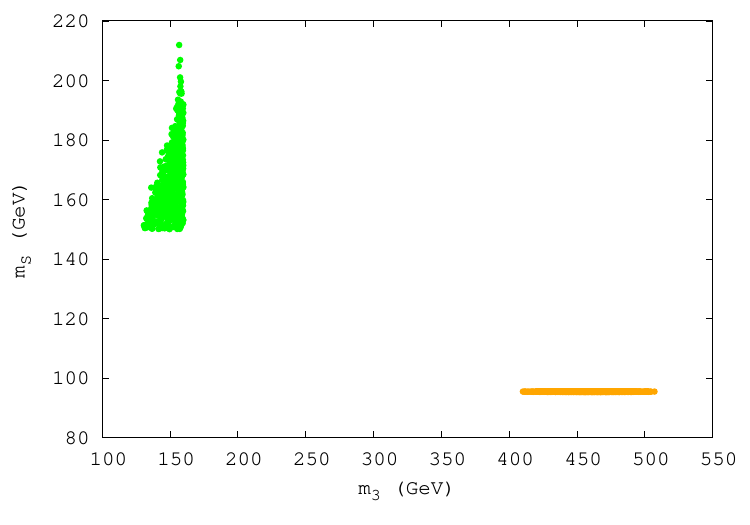} \ \ 
   \includegraphics[width= 7.5cm]{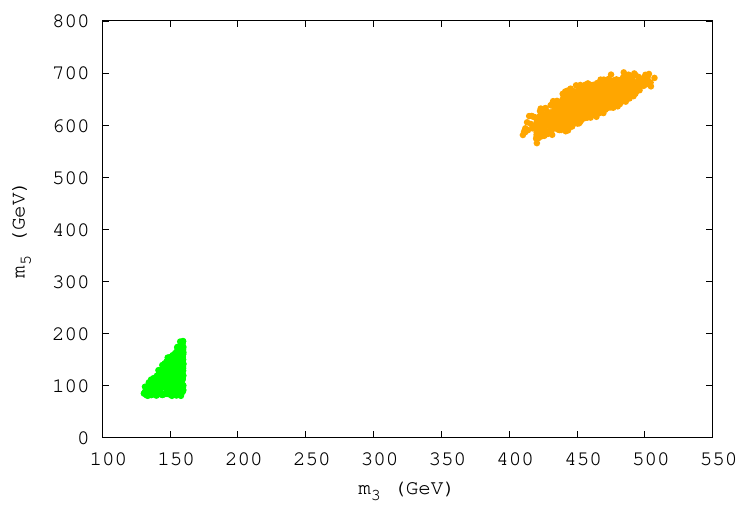} \\
  \includegraphics[width= 7.5cm]{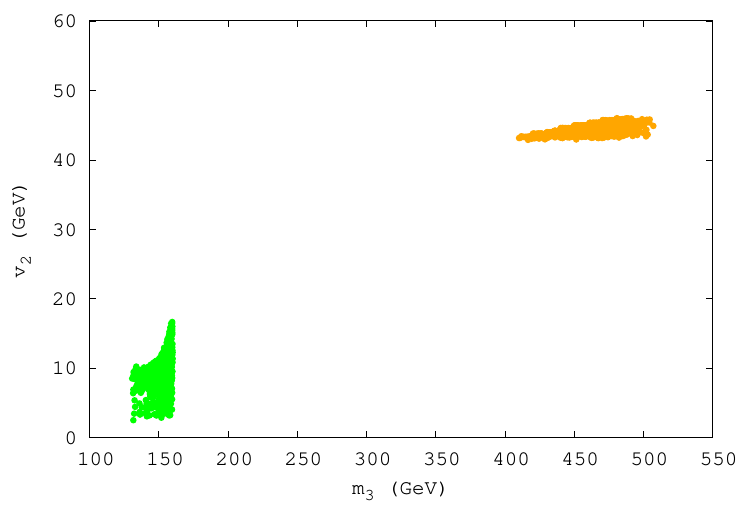} \ \ 
   \includegraphics[width= 7.5cm]{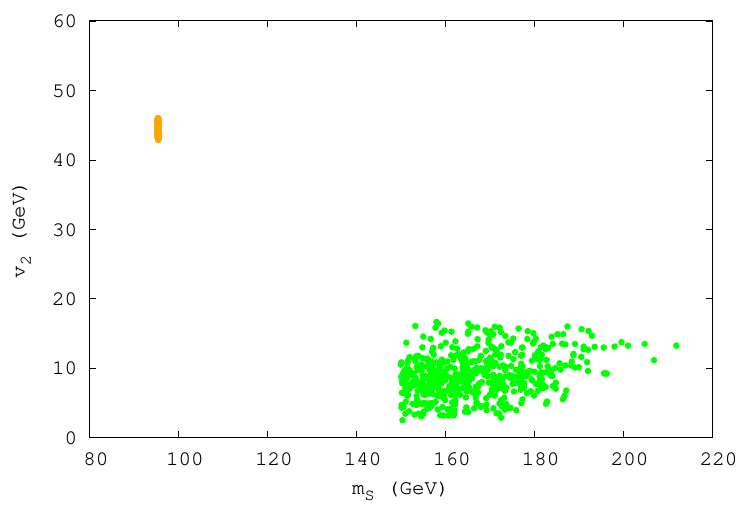} \\
  \end{center}
  \caption{\small Allowed parameter space in the $m_3-m_S$, $m_3-m_5$, $m_3-v_2$, $m_S-v_2$ plane. 
  All the coloured regions are allowed by the theoretical constraints, LHC Higgs signal data at $\sqrt{s}=13$ TeV. 
  The orange region is allowed for $m_S \approx 95$ GeV. 
  The green region is allowed for $m_3 \le 160$ GeV. 
  These two allowed regions are clearly mutually excluded.}
  \label{fig:overlapped}
  \end{figure}
 \section{Results}
 \label{Results}
This section shows the results from the parameter space scans for the low mass Higgs bosons. 
There are four masses ($m_{h,S,3,5}$) in the GM model as stated in the Section \ref{model}, out of which three are free parameters, as $m_h$ is always set to $125$ GeV. 
First, the mass of the SM-like Higgs boson is scanned in the range of $[95.355 : 95.444]$, and the other two masses $m_3$ and $m_5$ are scanned in the region $[80:1000]$. 
The triplet vev $v_2$ is varied in between $5 - 50$ GeV. 
The theoretical constraints together with the LHC Higgs signal strength data restrict the GM model parameter space to a great extent in this case. 
The scans result the allowed triplet vev $v_2$ over $42.5$ GeV, the mass $m_3$ over $400$ GeV, and the mass $m_5$ over $550$ GeV. 
The allowed regions for $m_S\,\approx \,95.4$ GeV are provided in the Fig. \ref{fig:overlapped} in orange shade. 
This clearly show that the presence of the $95$ GeV SM-like Higgs boson in the GM model disallow the presence of other low mass Higgs bosons. 
For completeness, the corresponding allowed parameter spaces for low mass charged Higgs boson $H_3^{\pm}$ ($m_3\, \le \, 160$ GeV) are also presented in the Fig. \ref{fig:overlapped} in the green shade. 
The mass $m_5$ is allowed below $200$ GeV in this case. 
The SM-like Higgs mass $m_S$ is below $220$ GeV here, and it is considered above $150$ GeV by choice. 
However, lowering the choice of $m_S$ does not allow the presence of $95$ GeV SM-like Higgs boson, as the allowed value of $v_2$ is higher than $42.5$ GeV for $95$ GeV SM-like Higgs boson, and the allowed value of $v_2$ is below $18$ GeV for the presence of light charged Higgs in the GM model. 
Therefore, one can easily conclude that the simultaneous presence of a $95$ GeV Higgs boson and a light charged Higgs boson in the GM model is not possible. 
%
 \section{Conclusions}
 \label{Conclusions}
%
CMS and ATLAS experiments of the LHC searched for the low mass SM-like Higgs boson with mass about $95$ GeV as well as charged Higgs boson with mass below the top quark mass. 
These searches encourage us to investigate the presence of both of these low mass Higgs bosons in different BSM models. 
The addition of a $SU(2)_L$ real and a $SU(2)_L$ complex triplet to the SM scalar sector delivers the scalar sector of the Georgi-Machacek model, such that, the custodial symmetry is reserved at tree level. 
The GM model consisting of a SM-like scalar beside the $125$ GeV Higgs boson and four singly charged scalars alongwith other four charged and neutral Higgs bosons persuades to explore whether low mass charged Higgs boson can be accommodated in this model together with low mass SM-like Higgs boson. 

The scalar sector of the GM model consists of two custodial singlets with different masses $m_{h,S}$ such that $m_h = 125$ GeV and $m_S$ can be greater than or less than $m_h$, one custodial triplet with three members of degenerate mass $m_3$, and one custodial quintuplet with five members of degenerate mass $m_5$. 
For the presence of a $95$ GeV SM-like resonance, $m_S = 95$ GeV, and one can find out that the triplet vev $v_2$ is above $40$ GeV when allowed by the theoretical constraints and the $13$ TeV LHC Higgs signal strengths. 
Again, the experimental data of LHC where the low mass charged Higgs decay into $\tau \nu_{\tau}$ show that the maximum value of $v_2$ cannot exceed $20$ GeV. 
Therefore, this paper shows that, the Georgi-Machacek model can only accommodate either the $95$ GeV Higgs boson with higher values of $m_{3,5}$ ($m_3 > 400$ GeV, $m_5 > 500$ GeV), or the low mass custodial triplet ($m_3 \le 160$ GeV) and quintuplet ($m_5 \le 200$ GeV) with the mass of the SM-like Higgs greater than $125$ GeV but less than $220$ GeV.

\vspace{0.5cm}
{\em{\bf Acknowledgements}} --- The author would like to 
acknowledge Department of Science and Technology, Government of India for financial support 
through ANRF-NPDF scholarship with grant 
no. PDF/2022/001784.


 
\end{document}